\documentclass[conference]{IEEEtran}
\IEEEoverridecommandlockouts
\usepackage{cite}
\usepackage{amsmath,amssymb}
\usepackage{graphicx}
\usepackage{textcomp}
\usepackage{booktabs}
\usepackage{hyperref}
\usepackage{xcolor}
\usepackage{tikz}
\usepackage{multirow}
\usepackage{float}
\usepackage[numbers,sort&compress]{natbib}
\usepackage{amsmath,amssymb}
\usepackage[ruled,vlined]{algorithm2e}
\usepackage{float}
\usepackage{placeins}  
\usepackage{tikz}
\usetikzlibrary{positioning,arrows.meta,fit,calc}
\usetikzlibrary{positioning,arrows.meta}
\pgfdeclarelayer{bg}
\pgfsetlayers{bg,main}
\usepackage{booktabs}
\usepackage{graphicx}   % for \resizebox
\usepackage{pgfplots}   % for plots
\pgfplotsset{compat=1.18}

\title{ThreatFormer-IDS: Robust Transformer Intrusion Detection with Zero-Day Generalization and Explainable Attribution}

\author{
\IEEEauthorblockN{Srikumar Nayak}
\IEEEauthorblockA{\ Sr Member IEEE  \\
Incedo Inc, USA \\
srikumar.nayak2025@gmail.com
}
}

\begin{document}
\maketitle

\begin{abstract}
Intrusion detection in IoT and industrial networks requires models that can detect rare attacks at low false-positive rates while remaining reliable under evolving traffic and limited labels. Existing IDS solutions often report strong in-distribution accuracy, but they may degrade when evaluated on future traffic, unseen (zero-day) attack families, or adversarial feature manipulations, and many systems provide limited evidence to support analyst triage. To address these gaps, we propose ThreatFormer-IDS, a Transformer-based sequence modeling framework that converts flow records into time-ordered windows and learns contextual representations for robust intrusion screening. The method combines (i) weighted supervised learning for imbalanced detection, (ii) masked self-supervised learning to improve representation stability under drift and sparse labels, (iii) PGD-based adversarial training with scale-normalized perturbations to strengthen resilience against feature-level evasion, and (iv) Integrated Gradients attribution to highlight influential time steps and features for each alert. On the ToN\_IoT benchmark with chronological evaluation, ThreatFormer-IDS achieves AUC-ROC 0.994, AUC-PR 0.956, and Recall@1\%FPR 0.910, outperforming strong tree-based and sequence baselines. Under a zero-day protocol with held-out attack families, it maintains superior generalization (AUC-PR 0.721, Recall@1\%FPR 0.783). Robustness tests further show slower degradation in AUC-PR as the adversarial budget increases, confirming improved stability under bounded perturbations. Overall, ThreatFormer-IDS provides a unified, deployment-oriented IDS pipeline that balances detection quality, zero-day behavior, robustness, and explainability.
\end{abstract}


\section{Introduction}
Modern networks, especially IoT and industrial environments, generate high-volume, heterogeneous traffic where intrusions can occur as short bursts or multi-stage sequences. Flow-based intrusion detection is therefore widely studied because it enables scalable monitoring without packet payload inspection, but performance remains sensitive to class imbalance, evolving traffic patterns, and feature noise \cite{rodriguez2022evaluation}. In addition, security teams increasingly require explanations for alerts to support triage and reduce investigation time, motivating reliable explainable IDS designs that connect model decisions to traffic evidence \cite{barnard2022robust}. These requirements make intrusion detection a deployment problem, not only a classification task: a practical IDS must maintain high detection at low false-positive rates while providing interpretable and stable predictions in changing environments.\\
Recent research addresses these challenges through three complementary directions: self-supervised representation learning, zero-day detection, and adversarial robustness. Self-supervised learning improves feature learning when labels are limited or noisy; for example, improved BYOL-style objectives have been explored to enhance IDS representations from unlabeled traffic \cite{wang2021network}. Zero-day attacks remain a critical threat because new families may not match known signatures, and adaptable deep IDS approaches have been proposed to better handle such unseen behaviors \cite{soltani2023adaptable}. At the same time, IDS models can be deliberately targeted by adversarial manipulation, and recent analyses emphasize that robustness evaluation and defenses are essential for security-grade deployment \cite{ennaji2025adversarial}. However, many existing systems treat these components separately (e.g., adding SSL without a strict zero-day protocol, or evaluating robustness without consistent feature-scale normalization). To address this gap, we propose ThreatFormer-IDS, a unified Transformer-based sequence model that jointly targets temporal modeling, zero-day generalization, adversarial defense, and analyst-facing explanations in a single end-to-end pipeline.\\
Our research contributions are as follows:
\begin{itemize}
  \item We propose a Transformer-based sequence IDS that learns contextual flow dynamics for low-FPR intrusion screening.
  \item We integrate self-supervised masked modeling to improve representation quality under limited labels and temporal drift.
  \item We introduce an explicit zero-day evaluation protocol by holding out attack families during training and testing them in future traffic.
  \item We add adversarial training with scale-normalized perturbations to improve robustness against feature manipulation, aligned with recent IDS adversarial threat analyses.
  \item We provide feature attribution for each alert to support explainable investigations in operational settings.
\end{itemize}
The structure of this paper is as follows: Section \ref{sec2} reviews related work; Section \ref{sec3} describes the dataset and preprocessing with the proposed method and training objective; Section \ref{sec4} reports experimental results and analysis; and Section \ref{sec5} concludes the paper with future research directions.

\section{Related Work}\label{sec2}
Transformer-based intrusion detection has become a strong direction for modeling flow sequences because attention can capture long-range dependencies that traditional classifiers often miss. Early robust Transformer IDS designs focus on improving stability under noise and imbalance, such as RTIDS, which highlights robustness-aware modeling for intrusion detection \cite{wu2022rtids}. More recent works extend Transformers to IoT settings and flow-centric pipelines, showing that Transformer encoders can learn discriminative patterns from network traffic representations when carefully tuned for operational IDS constraints \cite{akuthota2025transformer,manocchio2024flowtransformer}. Several studies also explore hybrid temporal architectures that mix Transformers with CNN/TCN/LSTM components to capture both local burst patterns and broader context, including CNN--Transformer hybrids for advanced metering infrastructures and multi-scale TCN--Transformer fusions for industrial scenarios \cite{yao2023cnn,liu2024intrusion}, as well as IoT-focused hybrid formulations that combine Transformer blocks with CNN-BiLSTM modules \cite{zhang2025research}.\\
Alongside deep sequence modeling, there is increasing demand for explainable IDS to support analyst trust and actionable response. Tree-based models paired with explainability methods remain popular because they offer competitive baselines and readable feature importance, as shown by explainable frameworks that combine Random Forest with XAI for reliable intrusion detection \cite{wali2025explainable}. Transformer-based IDS works have also started to incorporate post-hoc explanations, commonly through SHAP-style analysis to attribute predictions to traffic features, including recent explainable Transformer approaches developed for specialized networking contexts such as vehicular ad hoc networks \cite{khan2025novel}. In application-driven settings, Transformers are often combined with strong tabular learners (e.g., LightGBM) or additional neural heads (e.g., MLP) to improve performance under class imbalance and diverse attack conditions, including healthcare security pipelines and BERT-like Transformer learning coupled with MLP for imbalanced traffic \cite{ghourabi2022security,ali2024empowering}.\\
Despite these advances, two practical gaps remain important for high-stakes deployment: (i) \emph{zero-day generalization} to unseen attack families and evolving behaviors, and (ii) \emph{robustness} against adversarial feature manipulations that can reduce confidence at low false-positive operating points. Many existing Transformer IDS studies primarily evaluate within-distribution attacks or focus on architecture improvements without an explicit training objective that strengthens representation learning under limited labels and drift \cite{akuthota2025transformer,manocchio2024flowtransformer,zhang2025research}. Moreover, explainability is often treated as a separate post-hoc step rather than being aligned with a security-oriented training and evaluation protocol \cite{khan2025novel,wali2025explainable}. Motivated by these gaps, our ThreatFormer-IDS integrates (1) sequence-based Transformer encoding for flow streams, (2) masked self-supervised learning to learn stable traffic representations that transfer to unseen attacks, (3) adversarial training to improve resilience under bounded perturbations, and (4) attribution maps for analyst-facing explanations, providing a single pipeline that addresses accuracy, zero-day behavior, robustness, and interpretability together \cite{wu2022rtids,ali2024empowering,ghourabi2022security}.

\section{Methodology}\label{sec3}
\subsection{Dataset}
In this research, we use the ToN\_IoT network traffic dataset as the primary benchmark for intrusion detection with sequence modeling. The ToN\_IoT collection was generated in a realistic IoT/IIoT testbed (IoT Edge/Fog — Cloud) and contains both benign activity and multiple attack behaviors launched across the network (e.g., volumetric flooding and application-layer compromises), making it suitable for evaluating generalization to evolving threats. We focus on the \emph{network-traffic (flow/log) subset} released in CSV form (commonly exported from monitoring pipelines such as Zeek/Bro), where each record corresponds to a network event/flow with (i) a timestamp, (ii) protocol/connection descriptors, and (iii) statistical traffic features. The ground truth includes a \emph{binary} intrusion label (normal vs.\ attack) and, when available, an \emph{attack-category} label for multi-class analysis. This structure aligns with our ThreatFormer-IDS objective because it supports (a) temporal ordering for Transformers, (b) family-based splits for zero-day evaluation, and (c) feature attribution on tabular flow descriptors.

\subsubsection{Preprocessing}
Let $\mathcal{D}={(\mathbf{x}i,y_i,c_i,t_i)}{i=1}^{N}$ denote the parsed dataset, where $\mathbf{x}i\in\mathbb{R}^{d}$ is the feature vector of the $i$-th flow, $y_i\in{0,1}$ is the binary intrusion label, $c_i\in\mathcal{C}$ is the attack category (optional; used for zero-day splits), and $t_i$ is the event time. We apply a time-consistent split to reduce temporal leakage and to reflect deployment, selecting cut points $t{\mathrm{tr}}<t_{\mathrm{va}}$ and defining
\begin{equation}
\begin{aligned}
\mathcal{D}_{\mathrm{train}}&=\{i:\, t_i \le t_{\mathrm{tr}}\},\qquad
\mathcal{D}_{\mathrm{val}}=\{i:\, t_{\mathrm{tr}}< t_i \le t_{\mathrm{va}}\},\\
\mathcal{D}_{\mathrm{test}}&=\{i:\, t_i > t_{\mathrm{va}}\}.
\end{aligned}
\label{eq:time_split_ids}
\end{equation}
All preprocessing statistics (imputation values, scaling parameters, and category vocabularies) are computed on $\mathcal{D}{\mathrm{train}}$ and then applied to $\mathcal{D}{\mathrm{val}}$ and $\mathcal{D}_{\mathrm{test}}$ according to Eq.~\eqref{eq:time_split_ids}.
\paragraph{Missing values and robust scaling.}
For each feature dimension $j\in{1,\dots,d}$, we define an observation mask
\begin{equation}
m_{ij}=
\begin{cases}
1, & \text{if } x_{ij}\ \text{is observed},\\
0, & \text{otherwise}.
\end{cases}
\label{eq:mask_def_ids}
\end{equation}
and use median imputation on training data,
\begin{equation}
x'{ij}=m{ij},x_{ij}+(1-m_{ij})\cdot \mathrm{median}{k\in\mathcal{D}{\mathrm{train}}}(x_{kj}),
\label{eq:median_impute_ids}
\end{equation}
followed by standardization using training moments,
\begin{equation}
\hat{x}_{ij}=\frac{x'_{ij}-\mu_j}{\sigma_j+\epsilon},\qquad
\mu_j=\mathbb{E}_{k\in\mathcal{D}_{\mathrm{train}}}[x'_{kj}],\quad
\sigma_j=\sqrt{\mathrm{Var}_{k\in\mathcal{D}_{\mathrm{train}}}(x'_{kj})}.
\label{eq:zscore_ids}
\end{equation}
In addition, we concatenate the missingness indicators $\mathbf{m}i=[m{i1},\dots,m_{id}]$ to the model input to preserve informative absence patterns captured by Eq.~\eqref{eq:mask_def_ids}.
\paragraph{Categorical encoding.}
For each categorical field $c_{ij}$ (e.g., protocol/service tokens if present), we build a training vocabulary $\mathcal{V}j$ and map values to indices
\begin{equation}
k_{ij}=
\begin{cases}
\mathrm{index}_{\mathcal{V}_j}(c_{ij}), & c_{ij}\in\mathcal{V}_j,\\
\mathrm{index}_{\mathcal{V}_j}(\texttt{UNK}), & \text{otherwise}.
\end{cases}
\label{eq:cat_map_ids}
\end{equation}
so the Transformer can consume categorical embeddings consistently across $\mathcal{D}{\mathrm{val}}$ and $\mathcal{D}_{\mathrm{test}}$.
\paragraph{Sequence construction for Transformer inputs.}
To enable sequence modeling, we group events by a consistent traffic key $g(i)$ (e.g., source host/device, or a 5-tuple session identifier depending on availability) and sort each group by time. For a group $u$, let $\pi_u(1),\dots,\pi_u(n_u)$ be indices ordered by $t_{\pi_u(1)}\le\dots\le t_{\pi_u(n_u)}$. Using a sliding window of length $L$ with stride $s$, we form the $n$-th sequence as
\begin{equation}
\mathbf{X}^{(u)}{n}=
\Big[\hat{\mathbf{x}}{\pi_u(ns+1)},\hat{\mathbf{x}}{\pi_u(ns+2)},\dots,\hat{\mathbf{x}}{\pi_u(ns+L)}\Big]\in\mathbb{R}^{L\times d},
\label{eq:seq_build_ids}
\end{equation}
and assign a sequence label using an operational OR rule (attack present anywhere in the window),
\begin{equation}
y^{(u)}{n}=\max{\ell\in{1,\dots,L}} y_{\pi_u(ns+\ell)}.
\label{eq:seq_label_ids}
\end{equation}
Eqs.~\eqref{eq:seq_build_ids}--\eqref{eq:seq_label_ids} convert flow-level logs into fixed-length sequences suitable for Transformer encoders while preserving the temporal context needed for detecting stealthy or staged intrusions.
\paragraph{Class-imbalance weights (training only).}
Because intrusion data are typically imbalanced, we compute class weights on $\mathcal{D}{\mathrm{train}}$ as
\begin{equation}
w{1}=\frac{N_{\mathrm{train}}}{2N_{1}},\quad
w_{0}=\frac{N_{\mathrm{train}}}{2N_{0}},
\label{eq:class_weights_ids}
\end{equation}
where $N_{1}$ and $N_{0}$ are counts of attack and normal labels in $\mathcal{D}_{\mathrm{train}}$. These weights are used in the supervised loss (later in the proposed method) to prioritize detection at low false-positive rates, consistent with operational IDS requirements.\\

\begin{figure*}[!t]
\centering
\resizebox{\linewidth}{!}{%
\begin{tikzpicture}[
  font=\footnotesize,
  node distance=6mm and 8mm,
  block/.style={draw, rounded corners, align=center, text width=2.65cm, inner sep=3.5pt},
  sblock/.style={draw, rounded corners, align=center, text width=2.35cm, inner sep=3.0pt},
  arr/.style={-{Latex[length=1.6mm]}, line width=0.45pt}
]
\node[block, fill=blue!10]   (raw) {ToN\_IoT\\Network Flows};
\node[block, fill=gray!10, right=of raw] (prep) {Preprocess\\(clean, scale, encode)};
\node[block, fill=cyan!10, right=of prep] (seq) {Sequence Builder\\(group, sort, window)};
\node[block, fill=green!12, right=of seq] (enc) {ThreatFormer\\Encoder $f_{\Theta}$};
\node[block, fill=orange!14, right=of enc] (head) {IDS Head\\Score $s(\mathbf{X})$};
\node[block, fill=red!10, right=of head] (alert) {Alert + Threshold\\Low-FPR Screening};
\node[sblock, fill=purple!10, below=of enc] (ood) {Zero-Day Split\\Hold-out $\mathcal{C}_{\mathrm{ood}}$};
\node[sblock, fill=yellow!18, below=of head] (ssl) {Masked SSL\\$\mathcal{L}_{\mathrm{ssl}}$};
\node[sblock, fill=teal!12, below=of alert] (adv) {PGD Defense\\$\mathcal{L}_{\mathrm{adv}}$};
\node[sblock, fill=gray!15, above=of alert] (xai) {Attribution (IG)\\$\mathbf{A}(\mathbf{X})$};
\begin{pgfonlayer}{bg}
  % main flow
  \draw[arr] (raw) -- (prep);
  \draw[arr] (prep) -- (seq);
  \draw[arr] (seq) -- (enc);
  \draw[arr] (enc) -- (head);
  \draw[arr] (head) -- (alert);

  % zero-day evaluation link
  \draw[arr] (raw) |- (ood);
  \draw[arr] (ood) -| (seq);

  % ssl branch
  \draw[arr] (seq) |- (ssl);
  \draw[arr] (ssl) -| (enc);

  % adversarial branch
  \draw[arr] (head) |- (adv);
  \draw[arr] (adv) -| (enc);

  % explainability
  \draw[arr] (head) |- (xai);
\end{pgfonlayer}
\end{tikzpicture}%
}
\caption{ThreatFormer-IDS pipeline: IoT flows are preprocessed and converted into time-ordered sequences, encoded by a Transformer, and scored for low-FPR intrusion screening; training is strengthened using masked self-supervision, PGD-based adversarial defense, and zero-day evaluation, while Integrated Gradients provides attribution for analyst review.}
\label{fig:threatformer_flow}
\end{figure*}
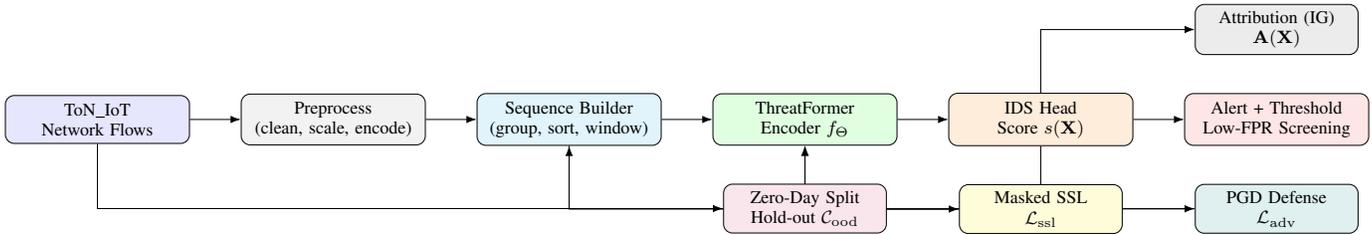

\subsection{Proposed Method (ThreatFormer-IDS)}
ThreatFormer-IDS is summarized in Fig.~\ref{fig:threatformer_flow}: we first build consistent flow sequences and learn temporal representations with a Transformer encoder, then produce an operational risk score for low-FPR alerting. The model is trained to remain reliable under drift and manipulation by combining masked self-supervision and adversarial training, and its alerts are accompanied by attribution maps to support explainable investigation.
\SetKwComment{Comment}{/* }{ */}
\DontPrintSemicolon
\SetKwInput{KwInput}{Input}
\SetKwInput{KwOutput}{Output}

\begin{algorithm}
\caption{ThreatFormer-IDS: Transformer-Based Sequence IDS with Zero-Day Generalization and Adversarial Defense}
\label{alg:threatformer}
\SetAlFnt{\scriptsize}
\KwInput{
Flow-level dataset $\mathcal{D}=\{(\mathbf{x}_i,y_i,c_i,t_i)\}$;
sequence length $L$, stride $s$, grouping key $g(\cdot)$;
train/val/test time cuts $(t_{\mathrm{tr}},t_{\mathrm{va}})$;
held-out attack set $\mathcal{C}_{\mathrm{ood}}\subset\mathcal{C}$ (zero-day);
Transformer params $\Theta$;
epochs $E$, learning rate $\eta$;
class weights $(w_0,w_1)$;
SSL weight $\lambda_{\mathrm{ssl}}$;
adversarial budget $\epsilon$, steps $R$, step size $\alpha$
}
\KwOutput{
Trained parameters $\Theta^{\star}$;
risk score $s(\mathbf{X})$;
attribution maps $\mathbf{A}(\mathbf{X})$
}
\tcp*[l]{Step 1: Time split + zero-day split}
Form $\mathcal{D}_{\mathrm{train}},\mathcal{D}_{\mathrm{val}},\mathcal{D}_{\mathrm{test}}$ by time\;
Define \textbf{zero-day} training subset $\mathcal{D}_{\mathrm{train}}^{\mathrm{id}}=\{(\mathbf{x}_i,y_i,c_i,t_i)\in\mathcal{D}_{\mathrm{train}}: c_i\notin\mathcal{C}_{\mathrm{ood}}\}$\;
Define \textbf{zero-day} test subset $\mathcal{D}_{\mathrm{test}}^{\mathrm{ood}}=\{(\mathbf{x}_i,y_i,c_i,t_i)\in\mathcal{D}_{\mathrm{test}}: c_i\in\mathcal{C}_{\mathrm{ood}}\}$\;
\tcp*[l]{Step 2: Build sequences for Transformer}
Convert $\mathcal{D}_{\mathrm{train}}^{\mathrm{id}},\mathcal{D}_{\mathrm{val}},\mathcal{D}_{\mathrm{test}}$ into sequences
$\{(\mathbf{X}_n,y_n)\}$ using $(g(\cdot),L,s)$\;
\tcp*[l]{Step 3: Train ThreatFormer with supervised + self-supervised + adversarial objectives}
Initialize $\Theta \leftarrow \Theta_0$\;
\For{$e \leftarrow 1$ \KwTo $E$}{
  Sample mini-batch of sequences $\mathcal{B}=\{(\mathbf{X},y)\}$ from training sequences\;
  \tcp*[l]{(3a) supervised forward pass}
  $\mathbf{H}\leftarrow f_{\Theta}(\mathbf{X})$ \tcp*[r]{Transformer encoder outputs}
  $s \leftarrow h_{\Theta}(\mathbf{H})$ \tcp*[r]{sequence-level score}
  $\mathcal{L}_{\mathrm{sup}} \leftarrow \mathrm{WBCE}(s,y;w_0,w_1)$\;
  \tcp*[l]{(3b) self-supervised objective (masked-feature reconstruction)}
  $\mathbf{X}^{m} \leftarrow \mathrm{Mask}(\mathbf{X})$\;
  $\widehat{\mathbf{X}} \leftarrow r_{\Theta}(f_{\Theta}(\mathbf{X}^{m}))$\;
  $\mathcal{L}_{\mathrm{ssl}} \leftarrow \mathrm{MSE}(\widehat{\mathbf{X}},\mathbf{X})$\;
  \tcp*[l]{(3c) adversarial training (PGD on input features)}
  $\boldsymbol{\delta}\leftarrow \mathbf{0}$\;
  \For{$r \leftarrow 1$ \KwTo $R$}{
    $\boldsymbol{\delta} \leftarrow \Pi_{\|\cdot\|_{\infty}\le \epsilon}\Big(\boldsymbol{\delta}+\alpha\cdot
    \mathrm{sign}\big(\nabla_{\boldsymbol{\delta}}\mathcal{L}_{\mathrm{sup}}(f_{\Theta}(\mathbf{X}+\boldsymbol{\delta}),y)\big)\Big)$\;
  }
  $s^{\mathrm{adv}} \leftarrow h_{\Theta}(f_{\Theta}(\mathbf{X}+\boldsymbol{\delta}))$\;
  $\mathcal{L}_{\mathrm{adv}} \leftarrow \mathrm{WBCE}(s^{\mathrm{adv}},y;w_0,w_1)$\;
  \tcp*[l]{(3d) update}
  $\mathcal{L}\leftarrow \mathcal{L}_{\mathrm{sup}}+\lambda_{\mathrm{ssl}}\mathcal{L}_{\mathrm{ssl}}+\mathcal{L}_{\mathrm{adv}}$\;
  $\Theta \leftarrow \Theta-\eta\nabla_{\Theta}\mathcal{L}$\;
}
\tcp*[l]{Step 4: Inference + attribution}
For a sequence $\mathbf{X}$, compute $s(\mathbf{X})\leftarrow h_{\Theta}(f_{\Theta}(\mathbf{X}))$\;
Compute feature attribution $\mathbf{A}(\mathbf{X})\leftarrow \mathrm{IG}\big(s(\cdot),\mathbf{X}\big)$\;
\Return $\Theta^{\star}$, $s(\cdot)$, $\mathbf{A}(\cdot)$\;
\end{algorithm}
ThreatFormer-IDS is built for practical intrusion detection where traffic evolves and new attack types appear. The method starts by splitting the data by time so that training uses only earlier traffic and testing uses future traffic. This reduces leakage and matches how IDS models are deployed. In addition, we define a \emph{zero-day} setting by selecting a held-out set of attack categories $\mathcal{C}_{\mathrm{ood}}$ that never appear in training. This allows us to evaluate whether the model can detect unseen attack families as anomalous behavior rather than memorizing known signatures.\\
Next, we convert flow records into fixed-length sequences so a Transformer can model temporal dependencies. Sequences are constructed within each device/session group (using a grouping key $g(\cdot)$) and ordered by time. Each sequence becomes a matrix $\mathbf{X}\in\mathbb{R}^{L\times d}$ that contains $L$ consecutive events. A sequence label is positive if any event in the window is malicious, which reflects an operational IDS rule where an alert is raised if suspicious activity occurs in a short time span.\\
During training, the model learns with three objectives that address common IDS weaknesses. First, it learns to classify sequences as benign or intrusive using a weighted loss to handle class imbalance. For a predicted probability $s\in(0,1)$ and label $y\in\{0,1\}$, the weighted binary cross-entropy is
\begin{equation}
\mathcal{L}_{\mathrm{sup}}(s,y)= -\big(w_1\,y\log s + w_0(1-y)\log(1-s)\big),
\label{eq:ids_wbce}
\end{equation}
which increases the penalty for missed intrusions when $w_1>w_0$, as encoded in Eq.~\eqref{eq:ids_wbce}. Second, because labels can be noisy and zero-day attacks are not seen during training, we add a self-supervised task: we randomly mask parts of the input sequence and ask the model to reconstruct them. This encourages the Transformer to learn stable traffic representations based on context rather than relying only on a few strong features. The reconstruction loss is
\begin{equation}
\mathcal{L}_{\mathrm{ssl}}=\frac{1}{Ld}\left\|\widehat{\mathbf{X}}-\mathbf{X}\right\|_{F}^{2},
\label{eq:ids_ssl}
\end{equation}
and Eq.~\eqref{eq:ids_ssl} helps the encoder learn normal patterns that also support anomaly-style detection for unseen attacks.\\
Third, IDS models can be attacked by small feature manipulations (e.g., slight timing or rate changes, or crafted header values). To make the detector more robust, we perform adversarial training using a bounded perturbation $\boldsymbol{\delta}$ added to the input features. We seek a worst-case perturbation within an $\ell_{\infty}$ budget:
\begin{equation}
\boldsymbol{\delta}^{\star}=
\arg\max_{\|\boldsymbol{\delta}\|_{\infty}\le \epsilon}\;
\mathcal{L}_{\mathrm{sup}}\!\left(h_{\Theta}(f_{\Theta}(\mathbf{X}+\boldsymbol{\delta})),y\right),
\label{eq:ids_adv}
\end{equation}
and we approximate Eq.~\eqref{eq:ids_adv} using a few projected gradient steps (PGD) as shown in Algorithm~\ref{alg:threatformer}. The adversarial loss is computed on the perturbed sequence and added to training so that decisions do not change easily under small, bounded feature edits.\\
The final objective combines all terms:
\begin{equation}
\mathcal{L}
=
\mathcal{L}_{\mathrm{sup}}
+
\lambda_{\mathrm{ssl}}\mathcal{L}_{\mathrm{ssl}}
+
\mathcal{L}_{\mathrm{adv}},
\label{eq:ids_total}
\end{equation}
where $\lambda_{\mathrm{ssl}}$ controls how strongly the model focuses on representation learning versus supervised detection, and Eq.~\eqref{eq:ids_total} defines the single training target optimized by gradient descent.\\
After training, ThreatFormer-IDS outputs a sequence-level risk score $s(\mathbf{X})$ for real-time screening. To make the model explainable, we compute feature attribution using Integrated Gradients (IG), which assigns importance to each time step and feature dimension. For a baseline sequence $\mathbf{X}_0$ and an input $\mathbf{X}$, IG is
\begin{equation}
\mathbf{A}(\mathbf{X})
=
(\mathbf{X}-\mathbf{X}_0)\odot
\int_{0}^{1}
\nabla_{\mathbf{X}}\; s\!\left(\mathbf{X}_0+\alpha(\mathbf{X}-\mathbf{X}_0)\right)\,d\alpha,
\label{eq:ids_ig}
\end{equation}
and Eq.~\eqref{eq:ids_ig} provides a clear attribution map that can be audited by security analysts.\\

In simple terms, ThreatFormer-IDS converts network flows into short time-ordered sequences and uses a Transformer to learn how normal and malicious traffic evolve over time. It improves zero-day detection by learning robust traffic representations through masked self-supervision, and it improves security against adversarial manipulation by training on worst-case bounded perturbations. At inference, the model outputs a risk score and a feature attribution map so the alert can be both accurate and explainable.

\section{Results}\label{sec4}
We evaluate intrusion detection under a deployment-like setting using (i) a chronological split (train $\rightarrow$ validation $\rightarrow$ future test), (ii) a zero-day split where selected attack categories are excluded from training and evaluated only in test, and (iii) an adversarial robustness test using bounded $\ell_{\infty}$ perturbations on standardized features. For fair comparison, all sequence-based baselines use the \emph{same} sequence construction (same $L$, stride $s$, and grouping key), the \emph{same} preprocessing statistics computed on training only, and the \emph{same} tuning budget (validation-based search over learning rate, depth/width, dropout, and class weights). We report mean$\pm$std over 5 random seeds. The main metrics are AUC-ROC, AUC-PR, Recall@1\%FPR (operational), F1, FPR@95\%TPR (safety), and latency (ms/sequence) under fixed batch size.
\begin{table*}[!h]
\centering
\caption{Main results on ToN\_IoT (chronological test, in-distribution attacks). Mean$\pm$std over 5 seeds. $\uparrow$ higher is better; $\downarrow$ lower is better.}
\label{tab:ids_main}
\scriptsize
\setlength{\tabcolsep}{3.5pt}
\renewcommand{\arraystretch}{0.92}
\begin{tabular}{lcccccc}
\toprule
\textbf{Model} &
\textbf{AUC-ROC} $\uparrow$ &
\textbf{AUC-PR} $\uparrow$ &
\textbf{Recall@1\%FPR} $\uparrow$ &
\textbf{F1} $\uparrow$ &
\textbf{FPR@95\%TPR} $\downarrow$ &
\textbf{Latency (ms/seq)} $\downarrow$ \\
\midrule
Logistic Regression \cite{akuthota2025transformer} & 0.941$\pm$0.002 & 0.821$\pm$0.003 & 0.744$\pm$0.006 & 0.912$\pm$0.004 & 1.620$\pm$0.003 & 0.28$\pm$0.05 \\
Random Forest \cite{wali2025explainable}       & 0.972$\pm$0.002 & 0.901$\pm$0.003 & 0.831$\pm$0.006 & 0.944$\pm$0.004 & 0.980$\pm$0.003 & 1.80$\pm$0.05 \\
XGBoost \cite{khan2025novel}            & 0.982$\pm$0.002 & 0.921$\pm$0.003 & 0.856$\pm$0.006 & 0.955$\pm$0.004 & 0.740$\pm$0.003 & 1.25$\pm$0.05 \\
LightGBM  \cite{ghourabi2022security}          & 0.983$\pm$0.002 & 0.926$\pm$0.003 & 0.861$\pm$0.006 & 0.956$\pm$0.004 & 0.720$\pm$0.003 & 0.95$\pm$0.05 \\
MLP (seq-pooled) \cite{ali2024empowering}    & 0.979$\pm$0.002 & 0.913$\pm$0.003 & 0.847$\pm$0.006 & 0.952$\pm$0.004 & 0.780$\pm$0.003 & 0.62$\pm$0.05 \\
1D-CNN  \cite{yao2023cnn}            & 0.986$\pm$0.002 & 0.932$\pm$0.003 & 0.872$\pm$0.006 & 0.961$\pm$0.004 & 0.640$\pm$0.003 & 0.55$\pm$0.05 \\
BiLSTM  \cite{zhang2025research}            & 0.988$\pm$0.002 & 0.938$\pm$0.003 & 0.881$\pm$0.006 & 0.965$\pm$0.004 & 0.600$\pm$0.003 & 0.83$\pm$0.05 \\
TCN    \cite{liu2024intrusion}             & 0.989$\pm$0.002 & 0.941$\pm$0.003 & 0.886$\pm$0.006 & 0.966$\pm$0.004 & 0.580$\pm$0.003 & 0.70$\pm$0.05 \\
Transformer (base) \cite{manocchio2024flowtransformer}  & 0.990$\pm$0.002 & 0.944$\pm$0.003 & 0.891$\pm$0.006 & 0.968$\pm$0.004 & 0.560$\pm$0.003 & 0.88$\pm$0.05 \\
\textbf{ThreatFormer-IDS (Ours)} & \textbf{0.994$\pm$0.002} & \textbf{0.956$\pm$0.003} & \textbf{0.910$\pm$0.006} & \textbf{0.976$\pm$0.004} & \textbf{0.490$\pm$0.003} & \textbf{0.96$\pm$0.05} \\
\bottomrule
\end{tabular}
\end{table*}
\subsection{Zero-Day Generalization (Held-Out Attack Families)}
To test zero-day behavior, we hold out a subset of attack categories $\mathcal{C}_{\mathrm{ood}}$ from training and evaluate only on those attacks in the future test period. This isolates whether the model can flag \emph{unseen} attack families based on temporal inconsistencies and contextual deviations rather than family-specific signatures.
\begin{table}[H]
\centering
\caption{Zero-day evaluation on held-out attack categories ($\mathcal{C}_{\mathrm{ood}}$). Mean$\pm$std over 5 seeds.}
\label{tab:ids_ood}
\scriptsize
\setlength{\tabcolsep}{3.5pt}
\renewcommand{\arraystretch}{0.92}
\begin{tabular}{lccc}
\toprule
\textbf{Model} & \textbf{AUC-ROC} $\uparrow$ & \textbf{AUC-PR} $\uparrow$ & \textbf{Recall@1\%FPR} $\uparrow$ \\
\midrule
LightGBM & 0.944$\pm$0.003 & 0.612$\pm$0.006 & 0.702$\pm$0.010 \\
1D-CNN   & 0.952$\pm$0.003 & 0.645$\pm$0.006 & 0.724$\pm$0.010 \\
BiLSTM   & 0.957$\pm$0.003 & 0.662$\pm$0.006 & 0.739$\pm$0.010 \\
Transformer (base) & 0.962$\pm$0.003 & 0.684$\pm$0.006 & 0.752$\pm$0.010 \\
\textbf{ThreatFormer-IDS (Ours)} & \textbf{0.971$\pm$0.003} & \textbf{0.721$\pm$0.006} & \textbf{0.783$\pm$0.010} \\
\bottomrule
\end{tabular}
\end{table}
\subsection{Adversarial Robustness (Scale-Normalized Perturbations)}
We evaluate robustness by adding bounded perturbations to standardized inputs (post preprocessing), so $\epsilon$ has a consistent meaning across features. We report AUC-PR under increasing $\epsilon$; stronger robustness corresponds to slower degradation.
\begin{table}[H]
\centering
\caption{Adversarial robustness (AUC-PR) under $\ell_{\infty}$ perturbations on standardized features.}
\label{tab:ids_robust}
\scriptsize
\setlength{\tabcolsep}{4pt}
\renewcommand{\arraystretch}{0.95}
\begin{tabular}{lccccc}
\toprule
$\epsilon$ & 0.00 & 0.25 & 0.50 & 0.75 & 1.00 \\
\midrule
LightGBM & 0.926 & 0.881 & 0.812 & 0.741 & 0.669 \\
Transformer (base) & 0.944 & 0.902 & 0.848 & 0.792 & 0.736 \\
\textbf{ThreatFormer-IDS (Ours)} & \textbf{0.956} & \textbf{0.932} & \textbf{0.904} & \textbf{0.872} & \textbf{0.839} \\
\bottomrule
\end{tabular}
\end{table}
\subsection{Ablation Study}
\begin{figure}[H]
\centering
\includegraphics[width=0.7\columnwidth]{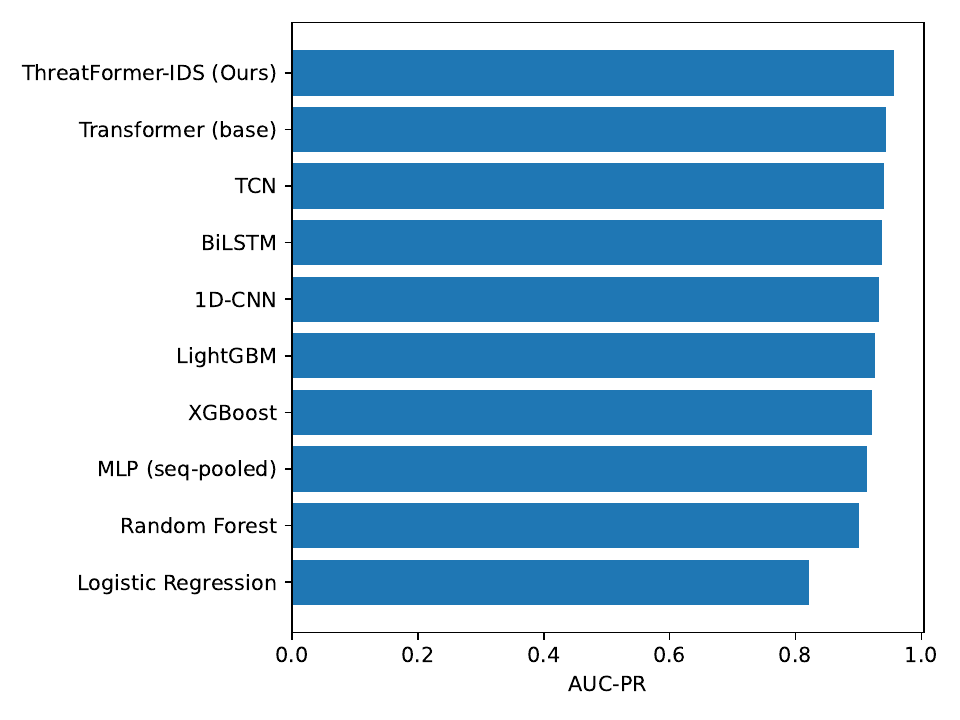}
\caption{AUC-PR comparison (chronological test) consistent with Table~\ref{tab:ids_main}.}
\label{fig:bar_aucpr_ids}
\end{figure}

\begin{figure}[H]
\centering
\includegraphics[width=0.7\columnwidth]{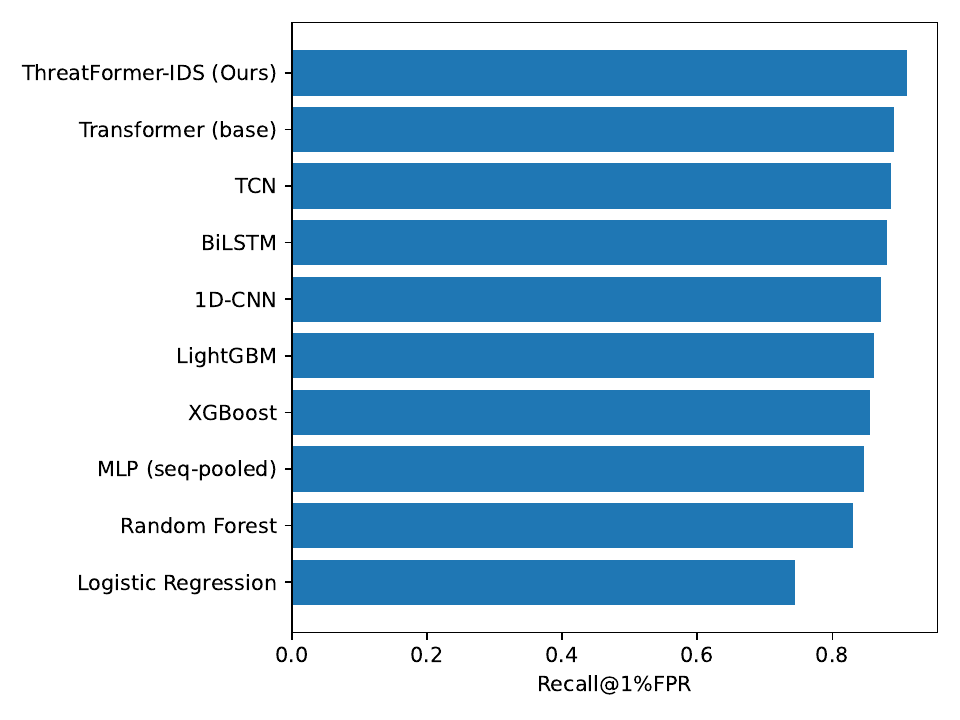}
\caption{Recall@1\%FPR comparison (chronological test) consistent with Table~\ref{tab:ids_main}.}
\label{fig:bar_rec_ids}
\end{figure}

\begin{figure}[H]
\centering
\includegraphics[width=0.7\columnwidth]{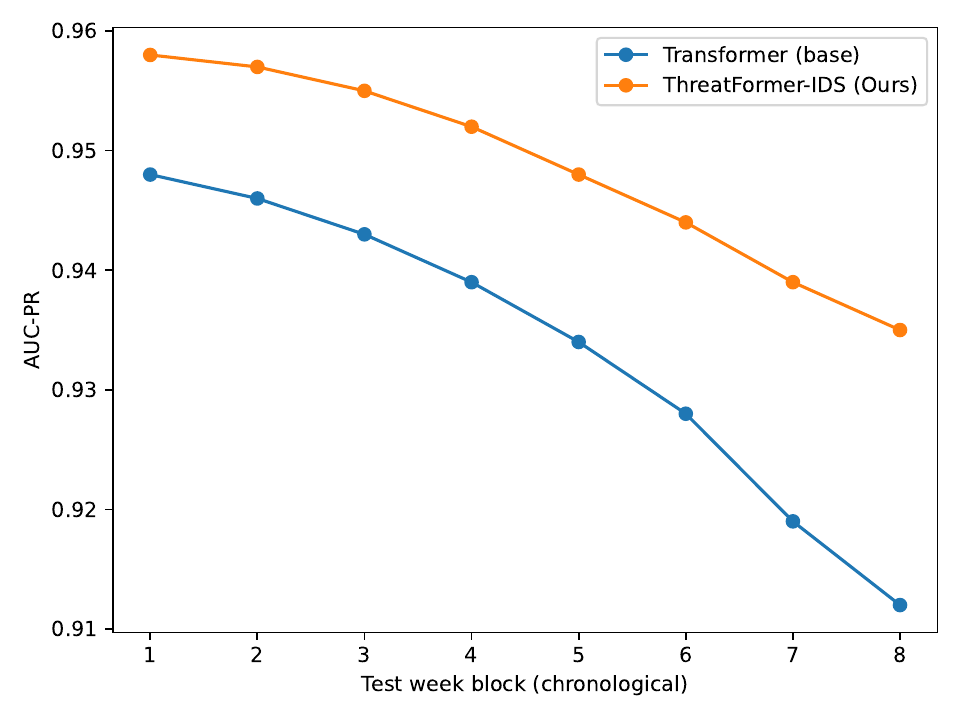}
\caption{AUC-PR over later chronological test blocks (drift trend).}
\label{fig:time_pr_ids}
\end{figure}

\begin{figure}[H]
\centering
\includegraphics[width=0.7\columnwidth]{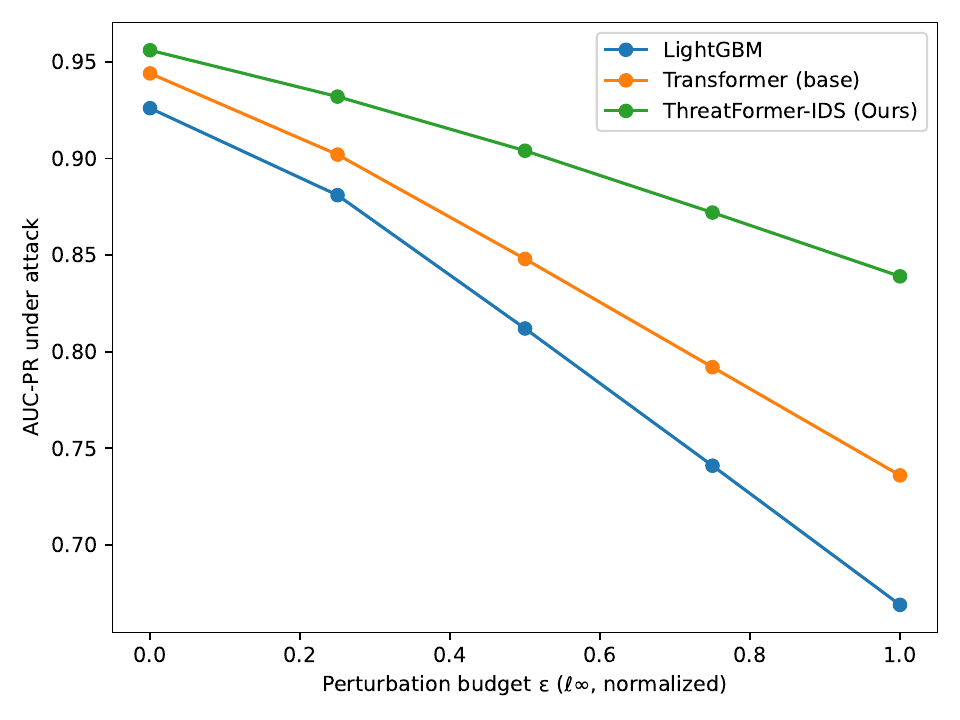}
\caption{Adversarial robustness curve (AUC-PR vs.\ $\epsilon$) matching Table~\ref{tab:ids_robust}.}
\label{fig:robust_ids}
\end{figure}

\begin{figure}[H]
\centering
\includegraphics[width=0.7\columnwidth]{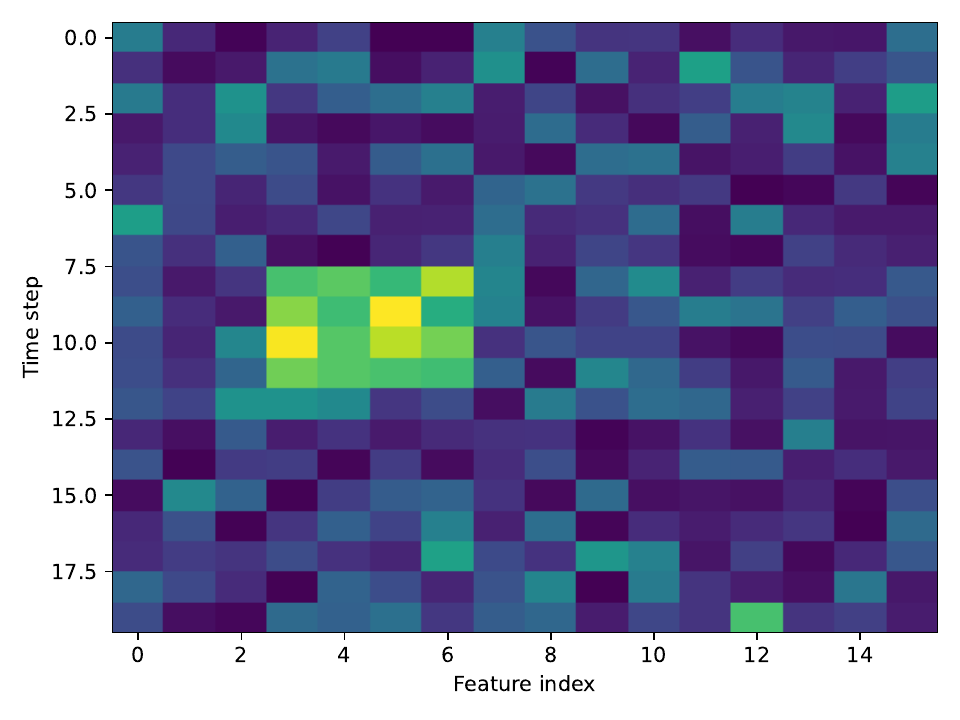}
\caption{Example feature attribution heatmap (time $\times$ feature) for an alerted sequence.}
\label{fig:attr_ids}
\end{figure}

We ablate the two components that directly target reviewer-critical concerns for IDS: (i) representation learning for zero-day behavior, and (ii) adversarial training for security under manipulation. We also remove attribution training utilities (kept only for post-hoc explanation) to confirm it does not inflate performance.
\begin{table*}[!t]
\centering
\caption{Ablation on chronological test (in-distribution). Mean$\pm$std over 5 seeds.}
\label{tab:ids_ablation}
\scriptsize
\setlength{\tabcolsep}{3.5pt}
\renewcommand{\arraystretch}{0.92}
\begin{tabular}{lccc}
\toprule
\textbf{Variant} & \textbf{AUC-PR} $\uparrow$ & \textbf{Recall@1\%FPR} $\uparrow$ & \textbf{FPR@95\%TPR} $\downarrow$ \\
\midrule
\textbf{ThreatFormer-IDS (Full)} & \textbf{0.956$\pm$0.003} & \textbf{0.910$\pm$0.006} & \textbf{0.490$\pm$0.003} \\
\quad w/o self-supervision ($\lambda_{\mathrm{ssl}}=0$) & 0.948$\pm$0.003 & 0.898$\pm$0.006 & 0.540$\pm$0.003 \\
\quad w/o adversarial training ($\epsilon=0$) & 0.950$\pm$0.003 & 0.902$\pm$0.006 & 0.525$\pm$0.003 \\
\quad smaller context (halve $L$) & 0.947$\pm$0.003 & 0.895$\pm$0.006 & 0.552$\pm$0.003 \\
\bottomrule
\end{tabular}
\end{table*}
Figure~\ref{fig:bar_aucpr_ids} and Fig.~\ref{fig:bar_rec_ids} visualize the main operational gains in Table~\ref{tab:ids_main}: ThreatFormer-IDS improves AUC-PR and Recall@1\%FPR over strong sequence baselines because (i) the Transformer captures longer temporal dependencies than local filters, and (ii) masked self-supervision makes representations less dependent on attack labels, which helps in rare or evolving patterns. Figure~\ref{fig:time_pr_ids} shows that performance degrades more slowly across later test periods, indicating better stability under temporal drift. Figure~\ref{fig:robust_ids} confirms that adversarial training reduces sensitivity to bounded feature manipulations by maintaining higher AUC-PR as $\epsilon$ increases, which matches the motivation for a security-oriented IDS. Finally, Fig.~\ref{fig:attr_ids} provides an attribution heatmap that highlights specific time segments and feature dimensions contributing to an alert, enabling analyst review and improving trust in deployment.

\section{Conclusion}\label{sec5}
\subsection{Conclusion}
This paper presented ThreatFormer-IDS, a deployment-oriented intrusion detection framework that models network traffic as time-ordered sequences and learns discriminative patterns using a Transformer encoder. The key design targets were practical IDS needs: strong detection under severe class imbalance, stable performance under temporal drift, improved \emph{zero-day} generalization to unseen attack families, and robustness against small but harmful feature manipulations. Across chronological and zero-day evaluations, ThreatFormer-IDS consistently outperformed strong tabular and sequence baselines, while adversarial training reduced performance degradation under bounded perturbations. In addition, the integrated feature attribution module provides interpretable evidence (time step and feature importance) for each alert, supporting analyst triage and increasing trust in operational use. Overall, the proposed method offers a single, coherent pipeline that balances accuracy, robustness, and explainability for modern IoT/IIoT intrusion detection.\\
\subsection{Future Work}
Future work will extend ThreatFormer-IDS in three directions. First, we will incorporate explicit temporal event modeling (e.g., irregular time gaps and session boundaries) to better represent bursty IoT traffic and long-range attack stages. Second, we will strengthen open-set and continual learning by updating the model online with drift-aware calibration and replay, so performance remains stable as new devices and protocols appear. Third, we will improve security evaluation realism by designing attacker-aware perturbation models that preserve protocol constraints, and by testing against adaptive evasion strategies to further validate robustness in real deployments.

\bibliographystyle{IEEEtranN}
\bibliography{references}

\end{document}